\documentclass[preprint,12pt]{elsarticle}

\usepackage{amssymb}
\usepackage{color}
\usepackage{amsmath}
\usepackage{caption}
\usepackage{subcaption}
\usepackage[noabbrev,capitalize]{cleveref}

\journal{Nonlinear Physics D}

\begin{document}

\begin{frontmatter}



\title{P\'eclet-number dependence of optimal mixing strategies identified using multiscale norms}


\author[inst1]{Conor Heffernan}

\address[inst1]{Department of Applied Mathematics and Theoretical Physics,
            Wilberforce Road,
            Cambridge,
            CB3 0WA,
            Cambridgeshire,
            United Kingdom}

\author[inst1,inst2]{Colm-cille P. Caulfield}

\address[inst2]{BP Institute,
            Bullard Laboratories,
            Madingley Road,
            Cambridge,
            CB3 0EZ,
            Cambridgeshire,
            United Kingdom}

\begin{abstract}
The optimization of  the mixing of a passive scalar at finite P\'eclet number $Pe=Uh/\kappa$ (where $U,h$ are characteristic velocity and length scales and $\kappa$ is the scalar diffusivity) is  relevant to many significant flow challenges across science and engineering. While much work has focused on identifying flow structures conducive to mixing for flows with various values of $Pe$,  there has been relatively little attention paid to how the underlying structure of initial scalar distribution affects the mixing achieved. In this study we focus on two problems of interest investigating this issue. Our methods employ a nonlinear direct-adjoint looping  (DAL) method to compute fluid velocity fields which optimize a multiscale norm (representing the `mixedness' of our scalar) at a finite target time.
First, we investigate how the structure of optimal initial velocity perturbations and the subsequent mixing changes between initially rectilinear `stripes' of scalar and disc-like `drops'. We find that the ensuing stirring of the initial velocity perturbations varies considerably depending on the geometry of the initial scalar distribution.
 Secondly, we examine the case of lattices of multiple initial `drops' of scalar and investigate how the structure of optimal perturbations varies with appropriately scaled P\'eclet number defined in terms of the drop scale rather than the domain scale. We find that the characteristic structure of the optimal initial velocity perturbation we observe for a single drop  is upheld as the number of drops and $Pe$ increase. However, the characteristic  vortex structure {\color{black} and associated mixing exhibits some nonlocal variability,} suggesting that  rescaling to a local $Pe$ {\color{black} will} not capture all the significant flow dynamics. 
\end{abstract}

\begin{highlights}
\item Direct-adjoint-looping method used to compute fluid flows to optimize mixing.

\item Questions of interest relate to the structure of the underlying passive scalar.

\item Qualitative differences between a rectilinear and disc geometry are found.

\item The disc geometry problem is scaled up and we find the base structure is preserved.
\end{highlights}

\begin{keyword}
keyword one \sep keyword two
\PACS 0000 \sep 1111
\MSC 0000 \sep 1111
\end{keyword}

\end{frontmatter}

\section{Introduction}
\label{sec:intro}
A robust understanding of fluid mixing is of huge importance in the study of various physical phenomena.
Irreversible scalar mixing is characterized by a relatively large scale `stirring' combined with smaller scale diffusion to homogenize an initially inhomogeneous distribution of a scalar field \cite{eckhart}.
One of the many difficulties of studying irreversible mixing  lies in the fact that there is no unified theory that rigorously defines it.
For example, an interesting question from a mathematical viewpoint is how exactly does one define the `mixedness' of a substance? Intuitively, the variance of a mean zero passive scalar $\theta$ on a domain $\Omega$, defined as
\begin{align}
    ||\theta(\cdot, t)||^2_{L^2} &= \frac{1}{\mu(\Omega)}\int_{\Omega}|\theta(x,t)|^2\,dx, \label{eq:vardef}
\end{align}
where $\mu$ denotes Lebesgue measure (area or volume) is a natural  measure of the mixedness of a scalar field. In the particularly simple situation when the domain is a 2D torus $\Omega = \mathbb{T}^2$, it can be shown  that \cref{eq:vardef} reduces to
\begin{align}
    ||\theta(\cdot, t)||^2_{L^2} &= \sum_{\mathbf{k}} |\hat{\theta}_{\mathbf{k}}|^2, \label{eq:vartorus}
\end{align}
where $\hat{\theta}_{\mathbf{k}}$ are the Fourier coefficients of $\theta$ and the sum is taken over all wavenumbers $\mathbf{k}$.

While this measure is intuitive, there are non-trivial issues which need to be addressed if the scalar field satisfies an advection-diffusion equation with appropriate boundary conditions and a divergence-free velocity field ${\mathbf{u}}$ defined with periodic boundary conditions with spatial period $L$ defined at points $(x,y)\in\Omega$

\begin{align}
\frac{\partial \theta}{\partial t} + \mathbf{u}\cdot\nabla\theta &= \frac{1}{Pe}\nabla^2\theta,\\
\nabla\cdot\mathbf{u} &= 0,\\
\theta(x,y,t) &= \theta(x+L,y,t),\\
\theta(x,y,t) &= \theta(x,y+L,t).
\end{align}

First in the limit when the diffusivity goes to zero, or equivalently when $Pe \rightarrow \infty$, the variance is actually a conserved quantity. Even when $Pe$ is finite yet large, the decay of the variance is generically expected to be relatively slow, which introduces computational challenges
to flow optimization calculations \cite{foures}.
A multiscale measure has been developed to overcome this problem, based around the use of Sobolev norms of negative index $-s$ \cite{mathew}:
\begin{align}
    ||\theta(\cdot, t)||^2_{H^{-s}} &= \sum_{\mathbf{k}\neq\mathbf{0}} |\mathbf{k}|^{-2s}|\hat{\theta}_{\mathbf{k}}|^2 . \label{eq:mixnorm}
\end{align}
This measure is similar to \cref{eq:vartorus} but weights wavenumbers differently (depending on $s$), with smaller wavenumbers (corresponding to larger scale structures) leading to larger values of this measure. This measure is commonly referred to as a `mix-norm' \cite{thiffeault_mix}, as if a scalar field is `mixed' (in the intuitive sense of having no large scale coherent structure) then the mix-norm is expected to be small.  A mix-norm is also advantageous as it is consistent with the mathematically rigorous ergodic mixing and any choice of $s>0$ is consistent with the theory \cite{lin}.

Mixing quantification is a necessary component for  mixing optimization subject to a  finite initial energy constraint. One successful method in particular has been to use a cost functional-based approach and maximize the energy growth at a target time $T$ \cite{aamo1,aamo2}. To solve this type of optimization problem a `direct-adjoint-looping' (DAL) method has been developed \cite{cherub,rich1,rich2}. The algorithm computes initial flows $\mathbf{u}$ which optimize a given cost functional at a prescribed target time $T$. It is advantageous in that it enforces the fully nonlinear Navier-Stokes equations as a constraint of the problem.
A benchmark of its success has been established by producing optimal perturbations which effectively homogenize a scalar field undergoing mixing. Interestingly, optimal perturbations
which minimize the mix-norm (with given index) at relatively short target times can be
excellent proxies for perturbations which minimize variance
at longer target times (and hence ensure thorough mixing). Furthermore,
 for a given initial energy, such mix-norm-minimizing perturbations are much
more efficient at mixing than perturbations
which are chosen to maximize (perturbation) energy growth
 \cite{foures,marc,verm1}. More recently, the DAL method has also been used to investigate how variations in the mix-norm index $s$  affects the ensuing mixing dynamics in a toroidal geometry of the associated `optimal' perturbations \cite{verm2,heff}.

The connections (and contrasts) between variance and mix-norm based strategies has been investigated for flows with passive and dynamic (i.e. the flow is density-stratified in a gravitational field) scalars \cite{foures,marc} as well as at high and low values of finite P\'eclet number in three-dimensional flows \cite{verm1}.  All these studies have demonstrated that the mix-norm is indeed a useful proxy for computing optimal mixing. A common theme of the previously considered mixing optimization problems has been to consider relatively simple rectilinear structures for the initial passive (or dynamic) scalar, typically with one or two interfaces between areas of high and low concentration, with $\theta\approx 1$ and $\theta\approx -1$ respectively such that the spatial mean is zero. Such considerations
were natural first steps in demonstrating that the DAL method can identify optimal perturbations, and also that the use of the mix-norm is appropriate and computationally efficient.

Now that we have confidence in the DAL method, in this paper we build on such previous studies \cite{foures,marc,verm1,verm2,heff} to investigate mixing optimization for
initial scalar distributions with more detailed structure. We wish to gain insight into the particular fluid-dynamical processes
which lead to mixing for different structures in the initial scalar distribution.  In particular, we attempt to answer two questions. First, we ask the question:

\noindent
\textit{Q1: what differences may be found in mixing results for rectilinear `stripes'  of scalar as compared to  disc-like `drops'?}.

To answer this question, we perform a comparative study between an initial scalar distribution of a rectilinear stripe  used previously \cite{verm2,heff}  and an initial scalar distribution of one disc-like drop. We find that there is a qualitative difference between flows
associated with these initial structures, with the stripe-like structure undergoing more mixing compared with the drop-like structure for the same initial perturbation energy. For both structures, the optimal initial perturbation takes the form of multiple vortices that localize along the interfaces (i.e. the locations where  $\theta$ vary rapidly) and hence enhance mixing. These vortices have a characteristic scale significantly smaller than both the stripe and the (single) drop.

Motivated by the generic observation of vortices localized at interfaces,  the  second question we ask is:

\noindent \textit{Q2: at what scale, if any, do these vortices cease to align along drop interfaces for smaller and smaller drops and if they do cease to appear what structures replace them?}

In other words,  if the initial structure is a lattice of sufficiently many and hence sufficiently small drops, does the initial optimizing velocity perturbation (in the form of interfacially-localized vortices) change from what has been found in \cite{foures,marc,verm1,verm2,heff} and if so, what sort of mixing dynamics and initial optimal perturbation(s) replace it?  The distribution and scales present in the mixing will be characterized by the number of drops in a (square) lattice which will be given as $4^{n-1}$ for $n\in\{1,2,...\}$. Indeed, an associated question is:

\noindent {Q2a: \it Is it possible to identify an appropriate self-similar rescaling  of the flow dynamics, in particular in terms of a local P\'eclet number in terms of the drop scale rather than the domain scale? }

As we are always considering the flow on a two-dimensional torus, the square lattices effectively allow us to consider whether  non-local mixing processes associated for example with broken symmetries in the mixing from drop to drop can actually lead to enhanced mixing.
To address these questions, the rest of the paper is organised as follows. In section \ref{sec:meth}, we briefly review the DAL method, and describe the various initial scalar distributions which we consider. We then present and interpret our results in section \ref{sec:res}, and finally draw our conclusions in section \ref{sec:conc}.


\section{Methods}
\label{sec:meth}
There are four key control parameters for the mixing optimization problems which we consider. Two are directly associated with the optimization problem:  the particular choice of the mix-norm index $s$ used in the objective functional and the target time  $T$ of the optimization problem over which the mix-norm is minimized.
The other two are associated with the properties of the flow:
the P\'eclet number $Pe$  and the Reynolds number $Re$, defined as
\begin{equation}
    Pe = \frac{U h}{\kappa},\
    Re = \frac{U h}{\nu},\label{eq:pedef}
\end{equation}
where $U, h$ are  characteristic velocity and length scales and $\nu, \kappa$ are the kinematic viscosity and scalar diffusivity respectively.
As we discuss further below,  when we consider initial
lattices of disc-like drops of scalar, care needs to be taken in the choice of the appropriate length scale $h$, given in terms of the individual drop rather than the domain. This also has implications
for the appropriate definition of the target time $T$, as $Pe$ may also be interpreted as the ratio of the characteristic
diffusive time scale $h^2/\kappa$ to the advective time scale $h/U$. We return to this key point below.

In \cite{foures,marc,verm1,heff}, the Schmidt number $Sc=\nu/\kappa$
 was kept fixed at unity, and so $Pe=Re$.  To assess the different behaviour of flows with either initial stripes or drops, and thus to answer the first question Q1 posed in the Introduction, we will
 allow $Pe$ to differ from $Re$.
 For the second question Q2, where we are interested in whether
 the mixing is inherently local or exhibits meaningful differences
 between single drops and lattices of drops (when appropriate rescaling
 of the key parameters is done), for simplicity we fix
 $Re=Pe$. We also fix the initial perturbation energy density, non-dimensionalized using the characteristic velocity scale $U$. Solutions of the optimization algorithm corresponding to these parameters will be denoted by $OA(s, T, Re, Pe)$, with subscripts to distinguish between the striped (SG) and disc (DG) initial scalar distributions. For comparison between the striped and disc geometry, we consider four different choices of $Re$ and $Pe$: $Re=100$ with  either $Pe=50$ or $Pe=500$;
  and $Pe=100$ with either $Re=50$ or $Re=500$. For either the single
  stripe or the single disc-like drop,  we also vary the index of the mix-norm $s= 0.5,1,2$. For the various values of mix-norm index, we choose the target time  $T = s$. \color{black}We make this choice due to our previous observation in \cite{heff} that increasing either $T$ or $s$ leads to a qualitatively similar change in both the structure of the optimal initial perturbation and  the subsequent time evolution of the vorticity distribution, in that increasing either $T$ or $s$ favours larger characteristic vortices in the initial
  optimal perturbation.\color{black} Therefore, keeping the ratio between these two control parameters leads to some simplification of the problem while still retaining the key dynamics of interest. When considering the drop lattice problem, we restrict attention to $s = 1, T=0.1,0.2,0.4,0.8$ and $Pe=400,800,1600,3200$.

We use the nonlinear DAL method \cite{rich2} to compute the initial velocity field which will minimize the value of the mix-norm at the prescribed target time $T$. The flow takes place in a 2D torus of length $2 \pi$ with $x$ and $y$ denoting the horizontal and vertical directions respectively. The velocity field $\mathbf{u}=(u,v)$ and pressure $p$ are governed by the incompressible Navier-Stokes equations and the passive scalar field $\theta$ is governed by a conventional advection-diffusion equation. Therefore, the non-dimensionalized equations governing the evolution of these variables are:

\begin{align}
    \frac{\partial \mathbf{u}}{\partial t} + \mathbf{u}\cdot \nabla \mathbf{u} &= -\nabla p + Re^{-1} \nabla^2 \mathbf{u}, \label{eq:ns}\\
    \nabla \cdot \mathbf{u} &= 0,\label{eq:div} \\
    \frac{\partial \theta }{\partial t} + \mathbf{u} \cdot {\nabla}\theta &= Pe^{-1} \nabla^2 \theta, \label{eq:ad}
\end{align}
where $Pe$ and $Re$ denote the P\'eclet number and Reynolds number respectively and are defined by Equation (\ref{eq:pedef}).

As the scalar field is passive,   we may solve  \cref{eq:ns,eq:div} and \cref{eq:ad} separately. In the case of the striped geometry, we choose \begin{align}
   \theta_{SG}(\mathbf{x},0) := \theta_0(\mathbf{x}) &= \tanh\left(6\left(x - \frac{\pi}{2}\right)\right) -  \tanh\left(6\left(x - \frac{3\pi}{2}\right)\right) - 1.
\end{align}
This corresponds to a smooth zero-mean scalar distribution with a vertical stripe, centred at $x=\pi$ of width  $\pi$ of positive $\theta \simeq 1$, bordered by stripes of negative $\theta \simeq -1$. For the disc `drop' geometry, we choose

\begin{align}
   \theta_{DG}^n(\mathbf{x},0) := \begin{cases}
   1, &\quad \text{ if } \mathbf{x}\in C_n \\
   -1, &\quad \text{ if } \mathbf{x}\not\in  C_n \label{eq:circ}
   \end{cases}
\end{align}
where $C_n$ denotes the set of $4^{n-1}$ discs with centres evenly distributed in the vertical and horizontal directions on a lattice and radius given by
\begin{align}
    r_n &= \sqrt{\frac{2\pi}{4^{n-1}}}.
\end{align}

In our study, we consider $n=1,2,3,4.$ We also require an initial condition for the velocity field $\mathbf{u_0} = \mathbf{u}(\mathbf{x}, 0)$ in order to solve the system. For the very first loop of the DAL method, we set $\mathbf{u_0}$  to be random noise. After each iteration of the loop, we update  $\mathbf{u_0}$ to give us our initial condition to evolve the system of \cref{eq:ns,eq:div,eq:ad}.

In order to identify the initial perturbation which optimizes the mixing of the initially inhomogeneous fluid, we seek to minimize an objective functional subject to a  constraint on the kinetic energy of the initial perturbation:
\begin{align}
    ||\mathbf{u_0}||^2_{L^2(\Omega)} &= 2 e_0 \mu(\Omega),
\end{align}
where $e_0=0.03$ is the non-dimensional perturbation energy density  and $\mu$ denotes the `volume' (i.e. the area) of the flow domain $\Omega$. \color{black}As has been previously demonstrated
in \cite{verm2,heff}, this perturbation energy density is chosen as an appropriate intermediate
value. \color{black}It is both sufficiently large to allow for the identification
of non-trivial initial flow structures, and yet sufficiently small so that mixing
is not too rapid and so there is still the possibility to distinguish the efficacy for mixing of different initial perturbation structures.

We define the objective functional as
\begin{align}
    \mathcal{J}(\theta (T)) = \frac{1}{2}||\theta(\mathbf{x}, T)||_{H^{-s}(\Omega)}^2 ,
\end{align}
i.e. (half) the value of the Sobolev norm of (negative) index $-s$ (which we refer to as the mix-norm of index $s$) at the target time $T$.
We may then define the constrained optimization problem of interest as \begin{align}\text{argmin } \mathcal{J}(\theta (T)) \text{ subject to } ||\mathbf{u_0}||^2_{L^2(\Omega)} = 2 e_0 \mu(\Omega),
\end{align}
where $\{\mathbf{u}, \theta\}$ solve the system of \cref{eq:ns,eq:div,eq:ad}. The initial condition $\mathbf{u}_0$  does not appear explicitly in the objective functional, but nevertheless it affects $\mathcal{J}$ through the evolution of the flow variables which are constrained by the system of \cref{eq:ns,eq:div,eq:ad}. These constraints are imposed of course by the use of appropriate Lagrange multipliers, the spatially and temporally evolving so-called \textit{adjoint} variables denoted by $\{\mathbf{u}^{\dagger}, p^{\dagger}, \theta^{\dagger}\} = \{\mathbf{v}, q, \eta\}$. This is explained in detail in (for example) \cite{verm2}  and we follow their approach here. We may define a Lagrangian as

\begin{align}
    \mathcal{L} &= \mathcal{J}(\theta (T)) - \Sigma_{I \in \{NS, AD, C, IC\}}\quad \mathcal{J}_I,
\end{align}
where
\begin{align}
    \mathcal{J_{NS}} &= \int_0^T\int_{\Omega} \mathbf{v} \cdot \left( \frac{\partial \mathbf{u}}{\partial t} + \mathbf{u}\cdot \nabla \mathbf{u} +\nabla p - Re^{-1} \nabla^2 \mathbf{u}\right), \\
    \mathcal{J_{AD}} &= \int_0^T\int_{\Omega} \eta \left(\frac{\partial \theta }{\partial t} + \mathbf{u} \cdot {\nabla} \theta - Pe^{-1} \nabla^2 \theta \right), \\
    \mathcal{J_{C}} &= \int_0^T\int_{\Omega} q \nabla \cdot \mathbf{u}, \\
    \mathcal{J_{IC}} &= \int_{\Omega} \mathbf{v_0} \cdot (\mathbf{u}(\mathbf{x}, 0) - \mathbf{u_0}).
\end{align}
Variation with respect to adjoint variables yields \cref{eq:ns,eq:div,eq:ad}. Similarly, variation with respect to the direct variables $\{\mathbf{u}, p, \theta\}$ results in the so-called `adjoint Navier-Stokes' equations
\begin{align}
    \frac{\partial \mathbf{v}}{\partial t} + \mathbf{u}\cdot \nabla \mathbf{v} &= -\nabla q - Re^{-1} \nabla^2 {\mathbf{v}} + \eta\nabla \theta,\label{eq:ans} \\
    \nabla \cdot \mathbf{v} &= 0, \label{eq:adiv}\\
    \frac{\partial \eta }{\partial t} + \mathbf{u} \cdot \nabla \eta &= - Pe^{-1} \nabla^2 \eta.\label{eq:aad}
\end{align}
At $t = T,0$ we also produce the following terminal and initial conditions
\begin{align}
    \mathbf{v}(\mathbf{x}, T) &= 0, \\
    \eta(\mathbf{x}, T) &= \sum_{\mathbf{k}\neq\mathbf{0}} |\mathbf{k}|^{-2s}Re\{ \hat{\theta}_{\mathbf{k}}(T)\exp{(i \mathbf{k}\cdot\mathbf{x})}\} , \\
    \mathbf{v_0} &= \mathbf{v}(\mathbf{x}, 0), \\
    \nabla_{\mathbf{u_0}} \mathcal{L} &= \mathbf{v_0},
\end{align}
where $\hat{\theta}_{\mathbf{k}}$ are the Fourier coefficients of $\theta$ and  $Re\{\cdot\}$ denotes taking the real part. Due to the negative diffusion terms $-Re^{-1}\nabla^2\mathbf{v}$ and $-Pe^{-1}\nabla^2\eta$, \cref{eq:ans,eq:adiv,eq:aad} must be integrated backwards in time to avoid numerical instability. These equations are then integrated backwards from $t = T$ to $t = 0$,  thus forming
a `direct-adjoint loop'.  Using a numerical technique from \cite{douglas} and with $\mathbf{u_0}^n$ and $\mathbf{v_0}^n$ denoting the direct and adjoint velocities at $t = 0$ after $n$ loops of this direct-adjoint-looping (DAL) method, the updated guess $\mathbf{u_{n+1}}$ can be calculated by $$\mathbf{u_0}^{n+1} = \cos(\phi) \mathbf{u_0}^n + \sin(\phi) \mathbf{w_0}^n,$$
where $\mathbf{w_0}^n$ denotes the scaled (by the energy constraint) adjoint velocity $\mathbf{v_0}^n$ projected onto the hypersurface tangential to the energy hypersphere at $\mathbf{u_0}^n$, as described in detail in \cite{foures2}. The angle of rotation $\phi$ is calculated by using a backtracking line search \cite{dennis}. This looping procedure is repeated until convergence has been reached as measured by the normalized residual $r$, defined by
$$r = \frac{||\nabla_{\mathbf{u_0}}\mathcal{L} ^{\perp}||_{L^2(\Omega)}^2}{||\nabla_{\mathbf{u_0}}\mathcal{L} ||_{L^2(\Omega)}^2},$$
where the symbol $\perp$ denotes projection onto the hyperplane tangential to the energy hypersurface. Since the energy is fixed, a small residual ($r \sim O(10^{-3})$) implies the gradient can only change by varying its magnitude which is not permissible due to the (explicitly imposed) energy constraint.

The direct and adjoint equations are solved with a $4^{th}$-order mixed Crank-Nicholson Runge-Kutta scheme with incompressibility enforced through a fractional step method \cite{foures,matlab}. Simulations were performed using a discretization of $N=256$ points in both the $x$ and $y$ directions \color{black}(we use $N = 512$ for the $Pe=3200$ calculation in \cref{sec:drop})\color{black}. For presenting the results graphically, the mix-norm and variance are scaled by the evolution of the purely diffusive passive scalar defined as
\begin{align}
    M_s(t) &= \frac{||\theta(\mathbf{x},t)||^2_{H^{-s}(\Omega)}}{||\theta_d(\mathbf{x},t)||^2_{H^{-s}(\Omega)}}, \label{eq:msdef}\\
    V(t) &= \frac{||\theta(\mathbf{x},t)||^2_{L^2(\Omega)}}{||\theta_d(\mathbf{x},t)||^2_{L^2(\Omega)}},\label{eq:vdef}
\end{align}
where $\theta_d$ is the solution of  \cref{eq:ad} but with the advective term dropped, and the dependence of the (scaled) mix-norm on the index $s$ is labelled by the subscript. Scaled in this way, it is possible to identify the extent to which fluid motions, and hence non-zero advection, affect the reduction in mix-norm and variance over time.


\section{Results}
\label{sec:res}
\subsection{Striped Geometry (SG) versus Disc Geometry (DG) Mixing}
\label{sec:comp}
In previous studies of nonlinear mix-norm optimization the Schmidt number is often kept constant and so $Re=Pe$. In \cite{heff}, we suggested varying both P\'eclet number and Reynolds number individually rather than keeping them equal, which we do here. Initially, we wish to address Q1, and so we wish to  investigate how the optimal perturbations change in structure and mixing performance depending on the geometry of the passive scalar's initial distribution.
\begin{figure}[h!]
\centering
\includegraphics[width=\textwidth]{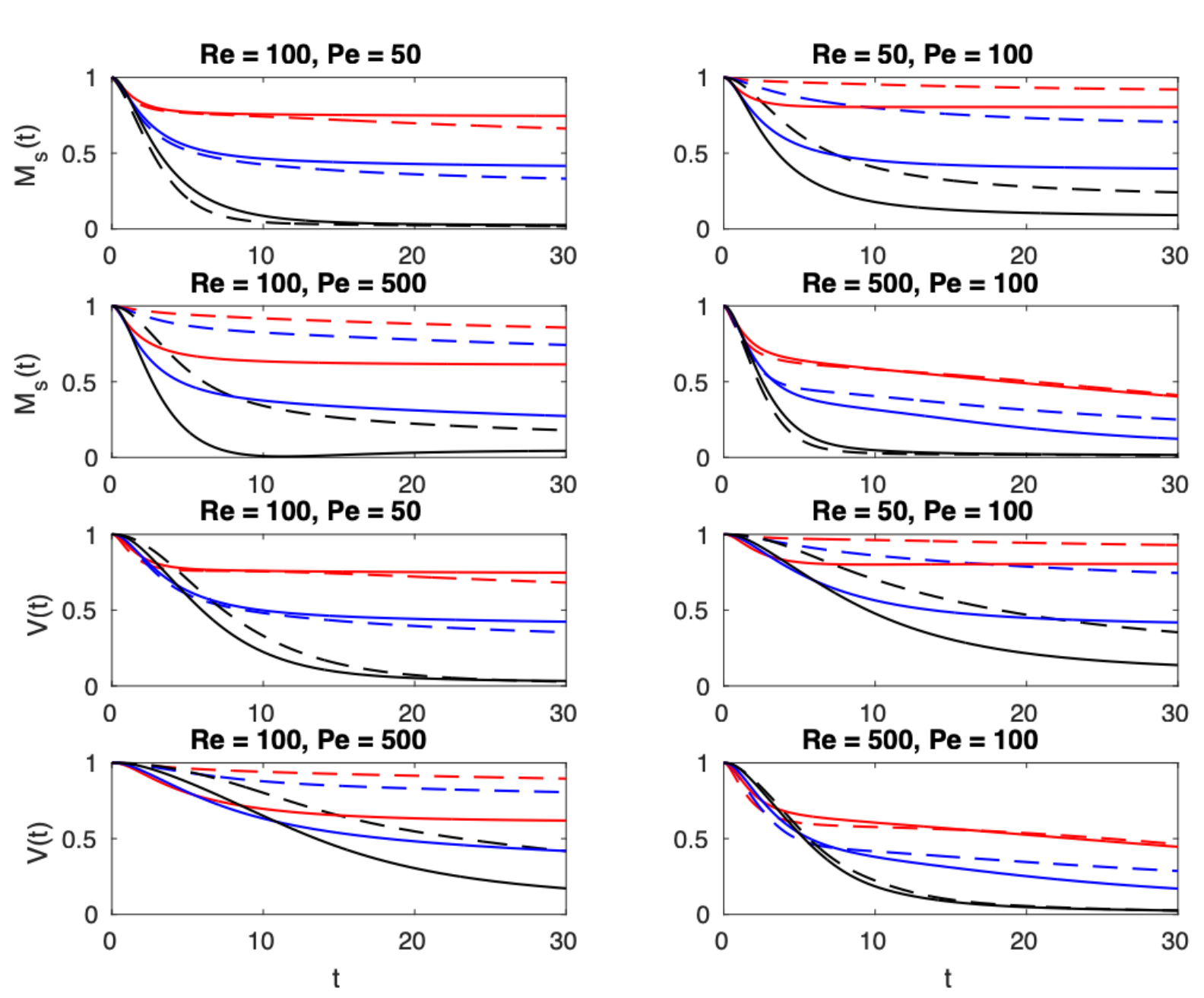}
\caption{Time evolution  of scaled mix-norm $M_s(t)$ and scaled variance $V_s(t)$, as defined in \cref{eq:msdef,eq:vdef}, for flows initially seeded with perturbations that minimize mix-norms (for a variety of values of index $s$) with $s=T=0.5$ shown in red, $s=T=1$ shown in blue, and $s=T=2$ shown in black. The solid lines denote the `SG' flows which are optimized to mix the striped geometry and the dashed lines denote the `DG' flows optimized to mix the disc geometry. Figures in the left-hand column correspond to flows with varying $Pe$ and fixed $Re=100$ while figures in the right-hand column correspond to flows with varying $Re$ and fixed $Pe=100$.}\label{fig:fig1}
\end{figure}
\begin{figure}[h!]
\centering
\includegraphics[width=1\textwidth]{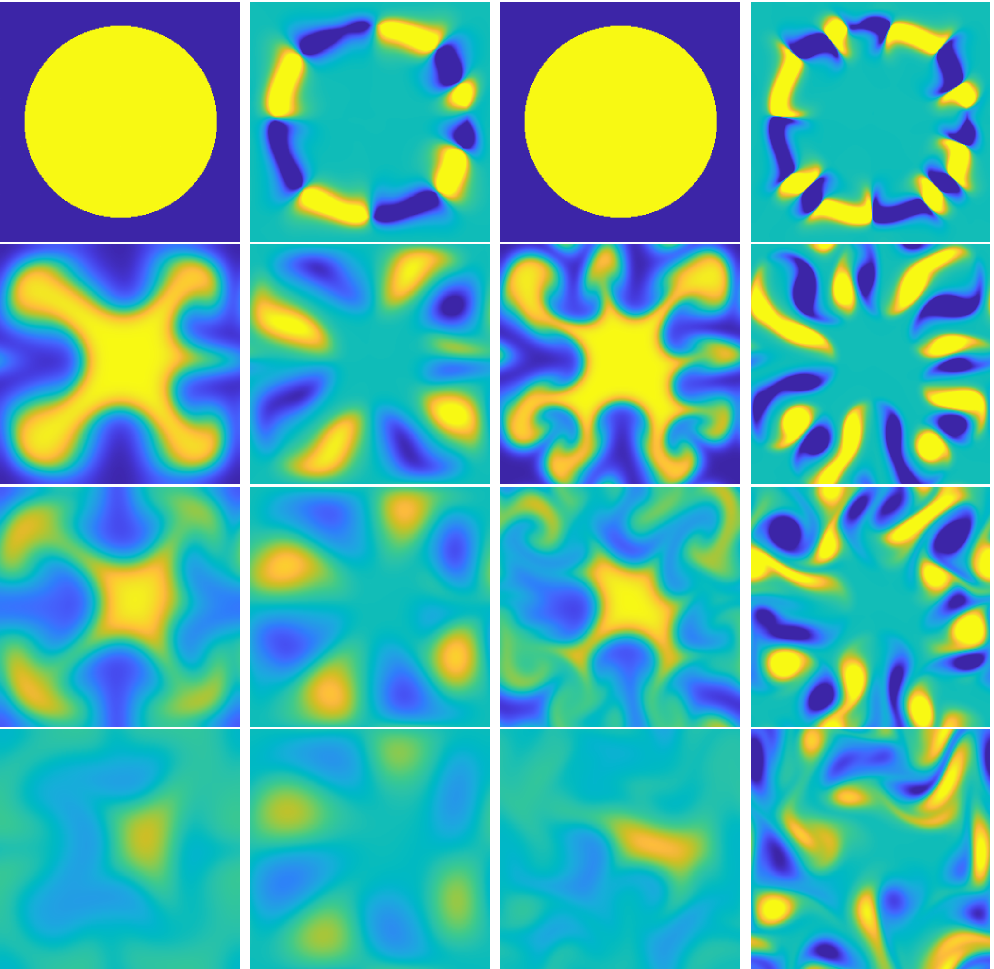}
\hspace*{1.2cm}
\includegraphics[width=0.6\textwidth]{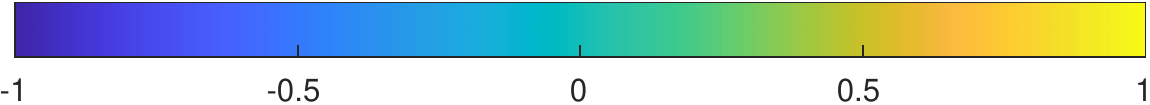}
\caption{Time evolution of passive scalar and vorticity for DG flows for $s=T=2$ and $Re=100,Pe=50$ (left columns) and $Re=500,Pe=100$ (right columns). Snapshots are taken at $t=0,5,10,20$.} \label{fig:fig7}
\end{figure}

We consider two such geometries here. The  first is the \textit{striped} geometry (SG) which consists of three layers, one on top of the other. The central stripe indicates the region where $\theta = 1$ and the two side stripes indicate $\theta=-1$. The second disc geometry (DG) we consider is a disc-like `drop' with radius $r=\sqrt{2\pi}$ (to ensure zero mean). The interior of the disc corresponds to $\theta=1$ and the surrounding region corresponds to $\theta=-1$.

\cref{fig:fig1} show the evolution of scaled mix-norm and variance respectively with solid lines denoting the striped geometry (SG) flows and dashed lines denoting the disc (or drop) geometry (DG) flows.
\begin{figure}[h!]
\centering
\includegraphics[width=1\textwidth]{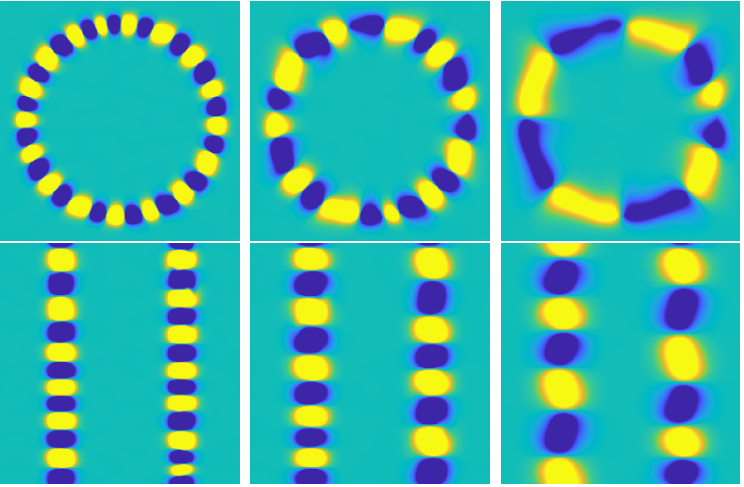}
\hspace*{1.2cm}
\includegraphics[width=0.6\textwidth]{colorbar-crop.pdf}
\caption{Initial vorticity distribution of the optimal
perturbations identified for mix-norm minimization for flows with $Pe=50$ and $Re\equiv100$. We keep $s=T$ and show $s=0.5$ (left column), $s=1$ (centre column) and $s=2$ (right column). Top row denotes DG and bottom row denotes SG.} \label{fig:fig3}
\end{figure}

For flows with varying $Pe$ and fixed $Re$ (left-hand column in \cref{fig:fig1}, there is  a qualitative difference between SG and DG flows. Similarly to the perturbations in \cite{heff}, the scaled mix-norms decay rapidly and then approach pure diffusion (and so the curves become close to  horizontal). The degree of decay  depends on $s$ and $T$ and also the initial scalar distribution. \color{black}SG flows (solid lines) typically exhibit a larger initial decay yet a smaller continued decay compared to the DG flows (dashed lines). \color{black}This behaviour is consistent with the case of varying $Re$ and fixed $Pe$ shown in the right-hand column. In both cases of $Re$, the more  rapid initial decay means that SG flows typically mix more thoroughly than DG flows at both target time and beyond.
\begin{figure}[h!]
\centering
\includegraphics[width=1\textwidth]{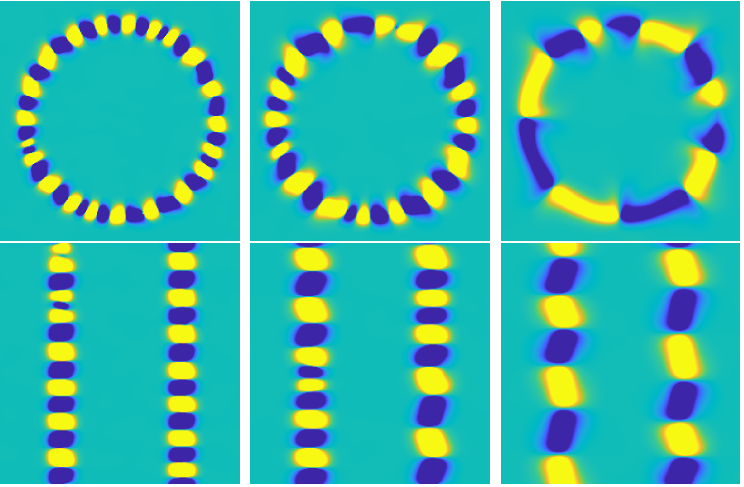}
\hspace*{1.2cm}
\includegraphics[width=0.6\textwidth]{colorbar-crop.pdf}
\caption{Initial vorticity distribution of the optimal
perturbations identified for mix-norm minimization for flows with $Pe=500$ and $Re\equiv100$. We keep $s=T$ and show $s=0.5$ (left column), $s=1$ (centre column) and $s=2$ (right column). Top row denotes DG and bottom row denotes SG.} \label{fig:fig4}
\end{figure}
\begin{figure}[h!]
\centering
\includegraphics[width=1\textwidth]{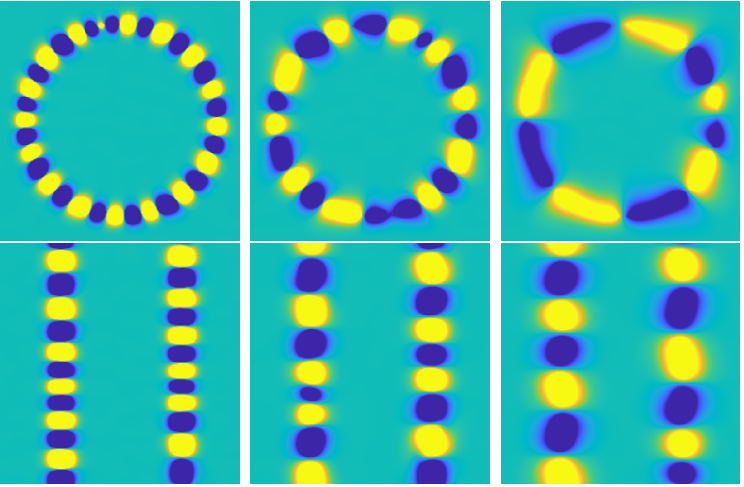}
\hspace*{1.2cm}
\includegraphics[width=0.6\textwidth]{colorbar-crop.pdf}
\caption{Initial vorticity distribution of the optimal
perturbations identified for mix-norm minimization for flows with $Re=50$ and $Pe\equiv100$.  We keep $s=T$ and show $s=0.5$ (left column), $s=1$ (centre column) and $s=2$ (right column). Top row denotes DG and bottom row denotes SG.} \label{fig:fig5}
\end{figure}

The qualitative difference between high and low P\'eclet number cases, conjectured in \cite{heff}, appears actually to be more dependent on variations in the Reynolds number  for SG flows (solid lines in \cref{fig:fig1}), as it is clear that
the difference between flows with $Pe=50$ and $Pe=500$ for $Re=100$ (left column) is not as marked as the difference between flows with $Re=50$ and $Re=500$ for $Pe=100$ . \color{black}However,
there does appear to be a more significant qualitative difference with $Pe$ for the DG flows.
The scaled variance for the  DG flows (dashed lines)  decays markedly more slowly than for the SG flows when $Pe=500$, whereas for the lower $Pe$ case both flows undergo a similar evolution.\color{black}
\begin{figure}[h!]
\centering
\includegraphics[width=1\textwidth]{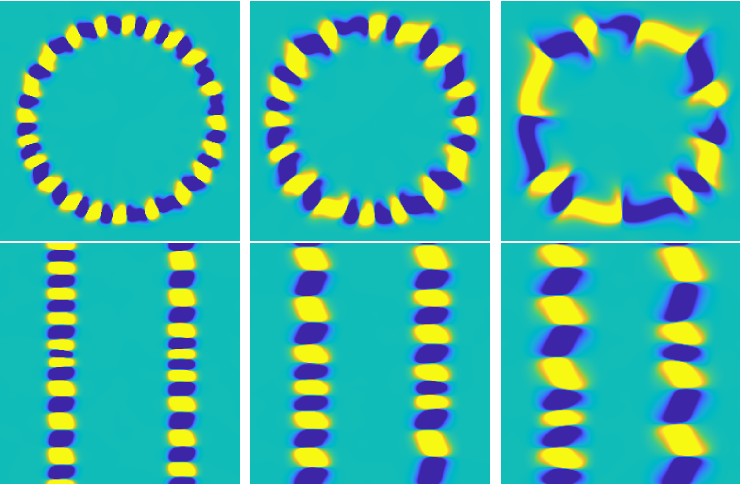}
\hspace*{1.2cm}
\includegraphics[width=0.6\textwidth]{colorbar-crop.pdf}
\caption{Initial vorticity distribution of the optimal
perturbations identified for mix-norm minimization for flows with $Re=500$ and $Pe\equiv100$.  We keep $s=T$ and show $s=0.5$ (left column), $s=1$ (centre column) and $s=2$ (right column). Top row denotes DG and bottom row denotes SG.} \label{fig:fig6}
\end{figure}

We now attempt to  understand the differences in mixing behaviours of  SG flows and DG flows, for the various control parameters we have considered. We first examine the differences between a low $Re$ and a high $Re$ DG flow. In \cref{fig:fig7}, we plot the time evolution of the vorticity and scalar fields for $OA_{DG}(2,2,100,500)$ and $OA_{DG}(2,2,50,100)$.
The initial structures consist of elongated vortices along the interface. The difference between these structures is that at higher $Re$ the vortices have perturbed corners similar to those observed in \cite{heff} at the same index and Reynolds number. We see from their evolution in \cref{fig:fig7} that the lower $Re$ field dissipates its energy more quickly; the vortices turn on their axis but retain a more rigid structure. However, in the case of the higher Reynolds number the vortices move freely throughout the domain with slower energy dissipation. Higher $Re$ flows more effectively mix the passive scalar as they are successful in fully tearing apart the disc whereas lower $Re$ flows do not do this as successfully and rely more on diffusion for full homogenization.
\color{black}As can be seen from the top rows of  \cref{fig:fig3,fig:fig4,fig:fig5,fig:fig6}, increasing target time $T$ increases the characteristic length scales of these vortices, and the perturbed corners are more strongly associated with higher $Re$ than with higher $Pe$. \color{black}  As identified in \cite{heff}, these perturbed vortices seem to lead to better mixing which also appears to hold here. This implies that at an intermediate value of mix-norm index $s\sim 1-2$ with high $Re$ this initial structure brings about a better mixture.

\begin{figure}[h!]
\centering
\includegraphics[width=1\textwidth]{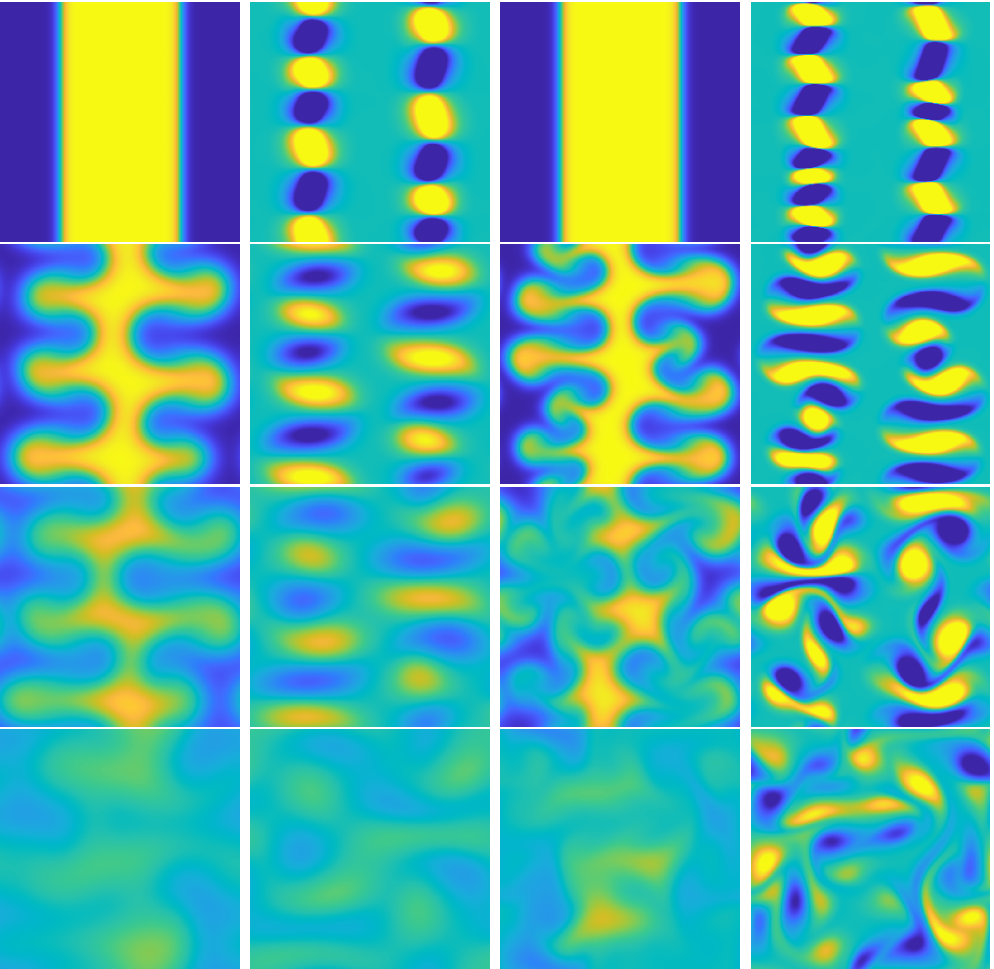}
\hspace*{1.2cm}
\includegraphics[width=0.6\textwidth]{colorbar-crop.pdf}
\caption{Time evolution of passive scalar (left) and vorticity (right) for SG for $s=T=2$ and $Re=100,Pe=50$ (left two columns) and $Re=500,Pe=100$ (right two columns). Snapshots are taken at $t=0,5,10,20$.} \label{fig:fig8}
\end{figure}

We now analyze the perturbations associated with the SG flows.
Analogously to above, in \cref{fig:fig8} we plot the time evolution of the vorticity and scalar fields for $OA_{SG}(2,2,100,50)$ and $OA_{SG}(2,2,500,100)$.
Again, the initial perturbation structure consists of vortices aligned along the interfaces of the three layers.  Similarly to the DG flow, we observe that at low $Re$  the initial vortices do not undergo much change in their position as the flow evolves in time. The scale set by the mix-norm manages to  break the central stripe effectively which leads to a good homogenization of the passive scalar by $t=20$. At higher $Re$ the perturbations manage to break the central stripe in the same time scale. However, the vortices are not fixed in position and cascade into different scales while moving around the domain. The energy dissipation in this case occurs over a longer timescale but both generate good mixing. Furthermore, as is apparent from the bottom rows of  \cref{fig:fig3,fig:fig4,fig:fig5,fig:fig6}, increasing target time $T$ again increases the characteristic length scales of these vortices. Also, increasing $Re$ once again appears
to lead to more changes in the  fine structure of these vortices than increasing $Pe$ does.

Fundamentally,  increasing $Re$ in both DG and SG flows causes a distortion in the otherwise smooth shape of vortices. This distortion then manifests itself by a cascade and free movement throughout the domain. A similar comparison can be made at low $Re$ between the geometries as the vortices stay fixed and dissipate energy more quickly. There is, however a distinct difference between the two classes of flow. As shown in \cref{fig:fig1}, it is clear that the stirring action in SG flows is superior to that which occurs in DG flows. A simple explanation may be framed in terms of the characteristic length scale ($d_c$) between regions where initially $\theta \simeq -1$ either side of the `central' region where $\theta \simeq 1$. For SG flows, $d_c \simeq \pi$, the width of the central `stripe' where $\theta=1$, while for (single drop) DG flows this length is nontrivially larger, as $d_c \simeq 2 r_1 = \sqrt{8 \pi}  \approx 5.0133$, the initial drop diameter.
As was observed in \cite{heff}, different parameter choices lead to different symmetry/asymmetry structure and this does appear to have a role in how the central region (with $\theta \simeq 1$) is distorted. However, the difference in how symmetry is characterized is not the same for SG flows and DG flows. Symmetry in SG flows is characterized by vortices on the rectilinear (vertical) interfaces between the different stripes, whereas for DG flows there is rotational symmetry around the edge of the central drop.

\subsection{The Drop Lattice Problem}
\label{sec:drop}
We now turn our attention solely to the DG flow, and  analyze the scenario where there is not one `drop' but many, arranged in a square lattice. The size and position of these drops will be given by \cref{eq:circ} (with mean zero). The problem of interest is to investigate the mixing dynamics as $n$ increases and investigate if increasing the number of drops leads to an appropriately self-similar scaled version of the previously considered DG flow with a single drop.

Periodic boundary conditions on one drop (as considered in \cref{sec:comp}) imply  an infinite lattice of drops next to one another with imposed symmetry. The question of interest is to investigate the behaviour if we increase the number of drops and decrease their size within the same $\Omega = [0, 2\pi]^2$ domain. We expect that as we scale up the number of drops we observe the same structure around the edge of the disc for each of the smaller counterparts (perhaps rotated around the interface). The central question is whether the optimization algorithm converges onto other solutions which diverge from the $n=1$ case, allowing for non-local interactions associated with broken symmetries between neighbouring drops and how the behaviour of these solutions compare qualitatively with one another. It is of interest to study how the potential introduction of finer scales set by the P\'eclet number can change the dynamics of the drop homogenization by comparing behaviour at different times, when the time is scaled appropriately, as discussed below.
\begin{figure}[h!]
\centering
\includegraphics[width=\textwidth]{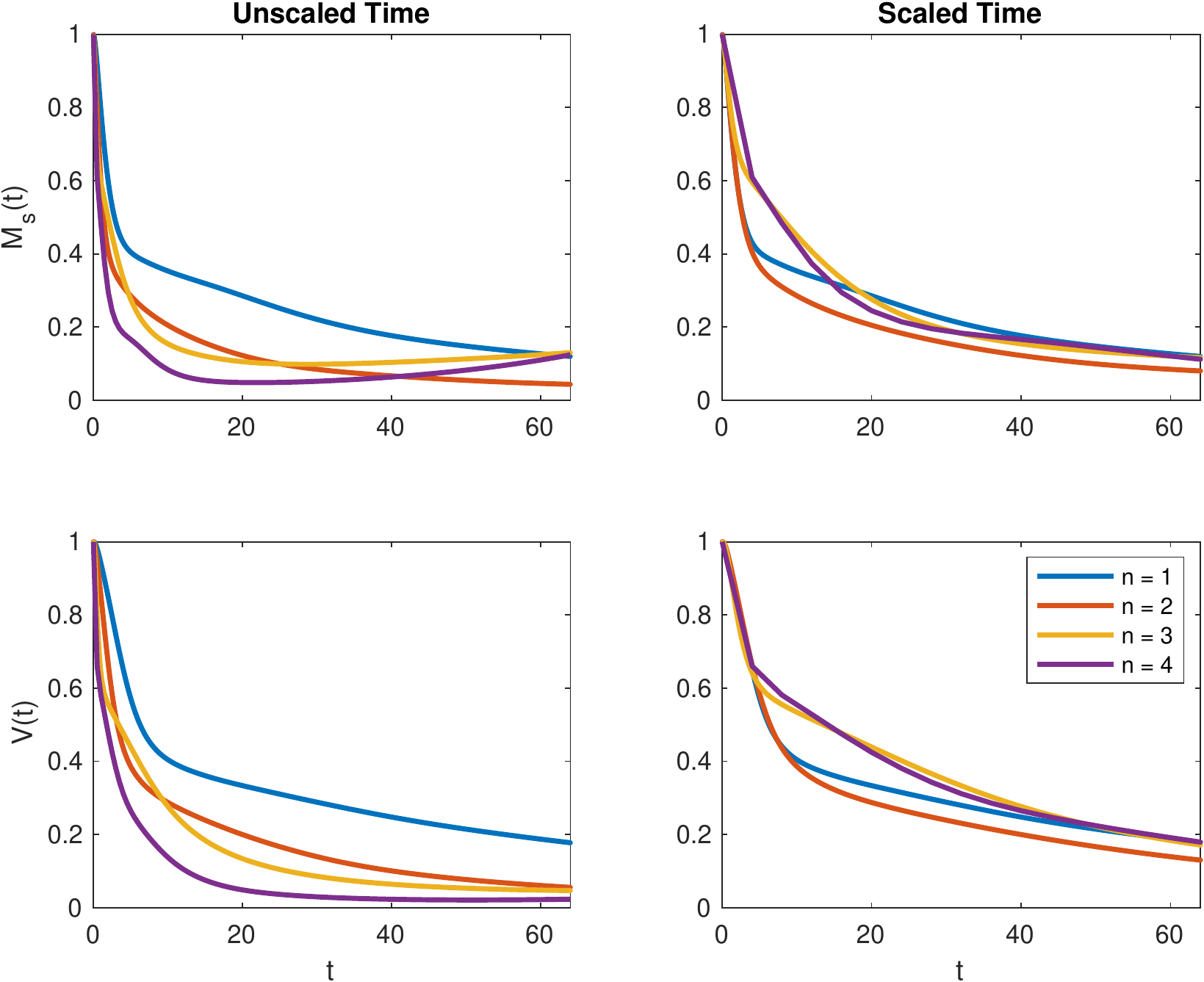}
\caption{Time evolution  of scaled mix-norm $M_s(t)$ and variance $V_s(t)$, as defined in \cref{eq:msdef} and \cref{eq:vdef} respectively, for flows initially seeded with perturbations that minimize mix-norms corresponding to $n=1,2,3,4$ for the drop lattice problem. } \label{fig:fig9}
\end{figure}
\begin{figure}[h!]
\centering
\includegraphics[width=0.75\textwidth]{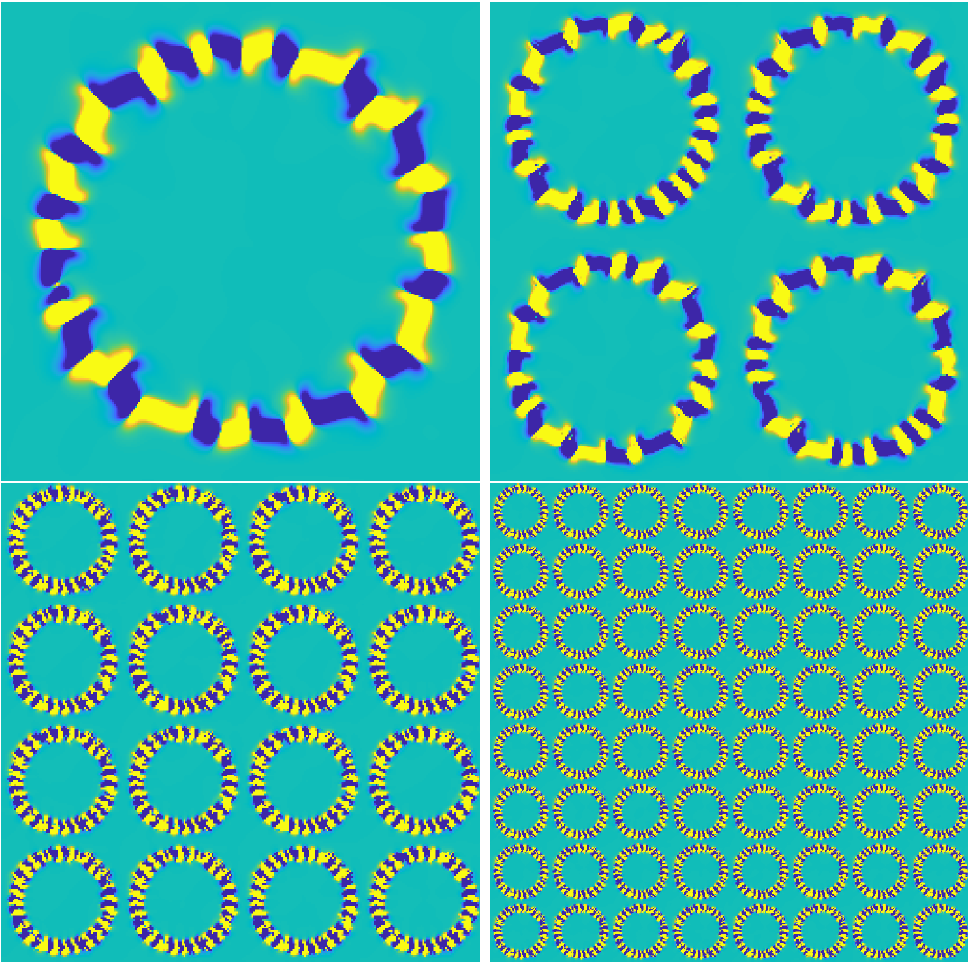}
\hspace*{1.2cm}
\includegraphics[width=0.6\textwidth]{colorbar-crop.pdf}
\caption{Initial vorticity distribution of the optimal
perturbations identified for mix-norm minimization for flows with   {\color{black} $4^{n-1}$ initial drops, $Pe=50 \times 2^{n-1}$ and target times $T_n=0.8 \times 2^{-(n-1)}$ for $n=1,2,3,4$.}} \label{fig:fig10}
\end{figure}

For this problem we fix $s = 1$ and we choose $T$ to scale appropriately (since we are changing scales).
For the flow with a single drop, the characteristic length scale $h$ used in the definitions of $Pe$ and $Re$ in \cref{eq:pedef} is appropriately related to the domain scale. However, as more drops are added, the appropriate length scale for a lattice of $4^{n-1}$ drops is {\color{black} $h_n=h/2^{n-1}$}. Assuming that the velocity scale is kept constant, (and that the physical properties of the fluid $\nu$ and $\kappa$ are also kept constant) we then expect the {\it local} P\'eclet number and Reynolds number for an individual disc also to be smaller by the factor $h_n/h=2^{-(n-1)},n=1,2,3,4$.

We set for a single drop $Pe=400$ and $T=0.8$ and then we increase and decrease respectively by a factor of $2$ as we increase from $n=1$ to $n=4$ (corresponding to 1 drop and 64 drops respectively). In problems of this type and those of \cite{foures,marc,verm1,verm2,heff,foures2}, the primary focus of study is the stirring action due to the optimal mixing strategies. Therefore, the decision to scale in this manner is to investigate dynamics on the local advective timescale $t_n = h_n/U$ rather than the diffusive timescale $t_n=h_n^2/\kappa$.

Underlying this rescaling is the assumption that the dynamics are entirely local, and that the key dynamics are dominated by advection.
Therefore, there are two ways in which there could be different behaviour even for the scaled cases. First,  there could be nonlocal effects due to broken symmetry for example which could modify the mixing dynamics. Secondly,  particularly at later times, when the flow is dominated by diffusive processes,
the diffusive time scale could become dominant. However, in this study, we choose to focus solely on the former.

\begin{figure}[h!]
\centering
\includegraphics[width=1\textwidth]{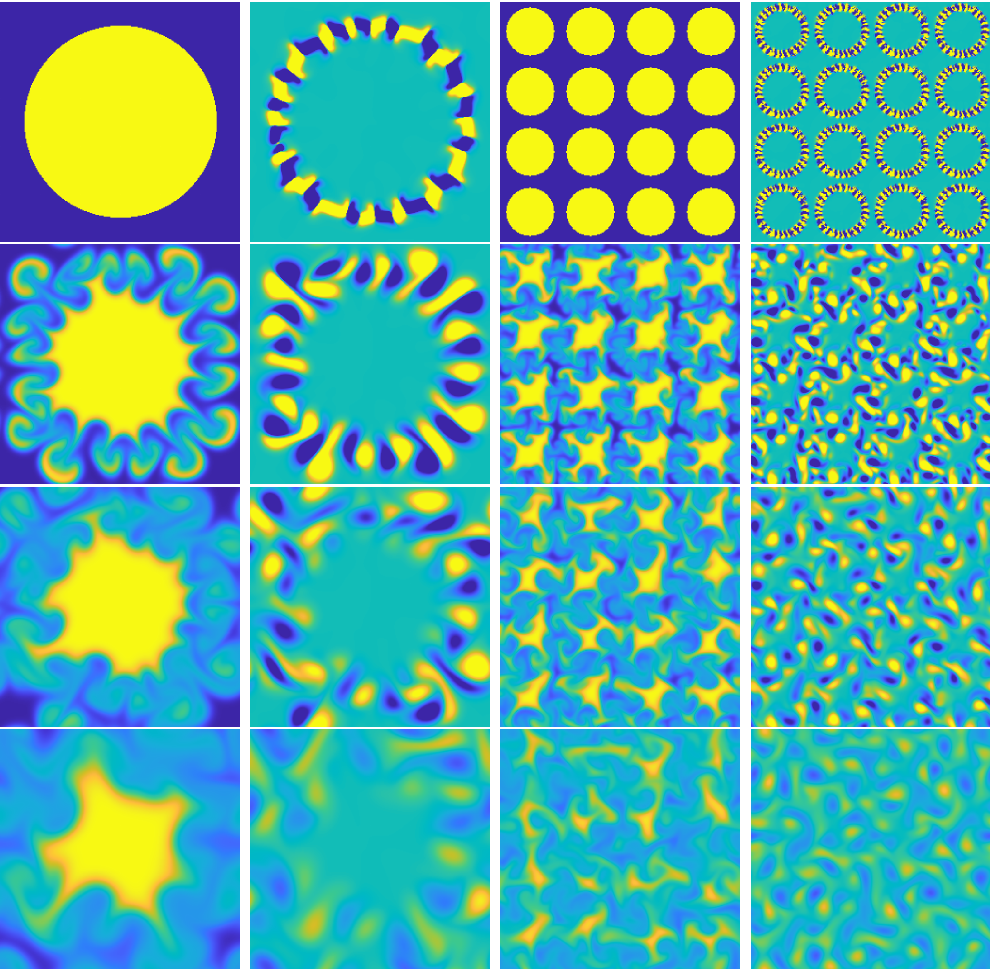}
\hspace*{1.2cm}
\includegraphics[width=0.6\textwidth]{colorbar-crop.pdf}
\caption{Time evolution of passive scalar (left) and vorticity (right)  for optimal perturbations that minimize the mix-norm with {\color{black} $s=1$ for  drop lattice problems with $Pe=200, n=2, T=0.4$ (left columns) and $Pe=800, n=3, T=0.2$} (right columns). Snapshots are taken at $t=0,5,10,20$.} \label{fig:fig11}
\end{figure}

Following this scaling approach, we choose $T=T_4=0.1$ for the $n=4$ case (with 64 drops) and increase by a factor of 2 as we go down through $n=3,2,1$ so that for the base case (with one drop) we have $T_1=0.8$. In the left column of \cref{fig:fig9} we plot the scaled mix-norm and variance for the evolution of the passive scalar with $n=1,2,3,4$ (shown in \cref{fig:fig10}) under the action of its corresponding optimal flow field. We also plot in the right column of \cref{fig:fig9} shifted mix-norm and variance with the time scale `stretched'. The behaviour we observe at the $Pe=400$ solution happens over a longer timescale than in the case of $Pe=3200$ and so we need to rescale in order to compare dynamics. As can be seen from \cref{fig:fig9}, there is a slight discrepancy in that the $n=4$ case undergoes a much quicker decay. An explanation of this lies in the examination of the vorticity of the solutions shown in \cref{fig:fig11}. We observe that as we increase the number of drops in our computational domain we do not get a repeat of the $n = 1$ structure. This change becomes clear passing through $n=2,3$ and is fully realized when $n=4$. \color{black}As can be seen from \cref{fig:fig11}, the $n=1$ case is asymmetric and disordered which results in a slower rate of decay. The $n=2$ case retains the corners observed for $n=1$ but this structure begins to break down into a more rigid structure as observed for $n=3,4$. After this transitionary phase, we oberve a symmetric solution for the $n=4$ case which leads to the most rapid decay. \color{black}  This would appear to suggest that the algorithm `catches' some smaller scale structures which cause a slow drift away from the $n=1$ case to something qualitatively different at $n = 4$. Indeed, the structure appears to diverge away from what our base case suggests.

The emergence of these finer scales leads to a more rapid homogenization (particularly at  $Pe = 3200$) and thus we observe that (after a rescale of time) we do not see the exact same mix-norm and variance decay between solutions. It appears that as we increase $Pe$, the optimization algorithm exploits the range of smaller scales now available as set by $Pe$ and uses them to perturb the base case in order to lead to a better mixture. Interestingly, these smaller scale structures are dominated by advective processes, as at later times the (rescaled) $n=4$ case is observed actually to homogenize more slowly, particularly relative to the purely diffusive behaviour (i.e. the denominator in \cref{eq:msdef}).
However, care needs to be taken not to over-interpret the late time behaviour of flows significantly after the imposed target time ($T=0.8$ on the rescaled plots), as it is only after that time that the $n=4$ case starts (apparently) to under-perform relative to the other cases.

We observe a very similar behaviour between solutions as P\'eclet number is increased. While there are some finer scales present as we increase $Pe$, they do appear to follow the same qualitative behaviour as the base case. We can see that as time progresses, the vorticity field deforms the interface and homogenizes the disc region into a thin stripe-like shape. This implies that the mixing dynamics behave in the same way as we scale up, even with the introduction of finer scales for higher $Pe$. Therefore, while there is some slight divergence away from the base case for lower $Pe$, the solution essentially behaves the same way as can be seen in the evolution of solutions in \cref{fig:fig11}. In fact, for higher $Pe$ (and higher $Re$ since Schmidt number is fixed at $Sc = 1$) we see that the algorithm scales up the base behaviour even as we approach a highly disordered regime. This suggests that as one exits an essentially laminar regime and finer scales are introduced, the underlying mixing dynamics still hold.

\begin{figure}[h!]
\centering
\includegraphics[width=\textwidth]{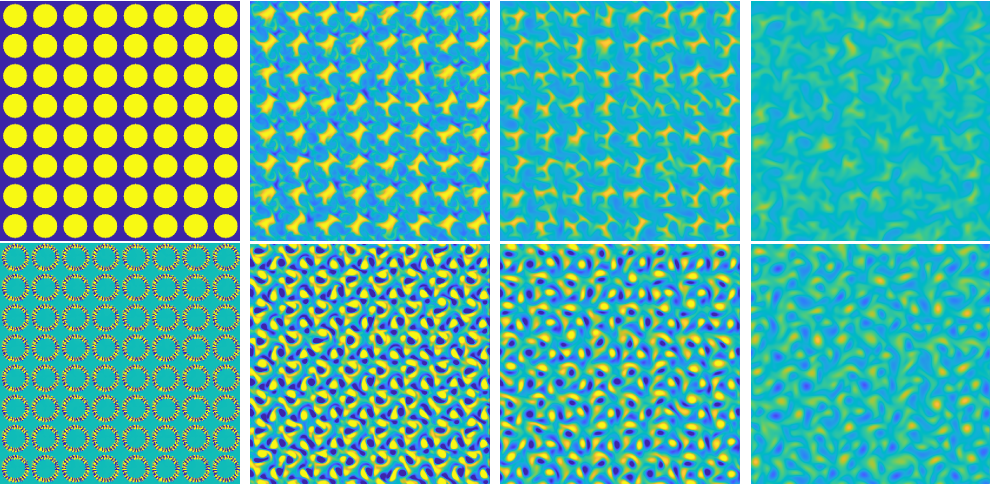}
\hspace*{1.2cm}
\includegraphics[width=0.6\textwidth]{colorbar-crop.pdf}
\caption{Time evolution of passive scalar (top row) and vorticity (bottom row)  for the drop lattice problem with {\color{black} $Pe=3200, n=4, s=1, T=0.1$.} Snapshots are taken at $t=0,5,10,20$.} \label{fig:fig14}
\end{figure}

To conclude this section, we comment on the emergence of a symmetric structure in the drop lattice at $Pe=3200$. In \cref{fig:fig14} we show the evolution of the passive scalar and vorticity. The evolution shows that the structure is perfectly symmetric, with each local drop deforming in the same way. This is in stark contrast to the lower $Pe$ cases which stretches and folds the drops asymmetrically but maintains essentially the same qualitative homogenization as shown in \cref{fig:fig11}. It appears that as we scale up, the finer scales tend to impose a rigid structure on the drop lattice. We conjecture that as $Pe$ increases, the optimal mixing field favours a rigid symmetric structure. However, if we compare this with the $Pe=800$ case, for example, we observe that the lower number of drops present leads to an asymmetric lattice of vortices. This leads to disordered mixing dynamics but as we can see from \cref{fig:fig9}, this asymmetry both at unscaled and rescaled time decays more rapidly than in the $Pe=3200$ solution which suggests that the symmetry may not necessarily be preferable for optimal mixing.

\section{Conclusion}
\label{sec:conc}
In this paper, we have studied two problems relating to the P\'eclet number dependence of optimal mixing strategies. These problems were studied using the nonlinear direct-adjoint-looping method \cite{rich1,rich2} to compute optimal flows which minimize the mix-norm at a prescribed target time. The central theme of both of the problems investigated concerned how optimal mixing relates to the structure of the underlying passive scalar.

In the first problem, we compared optimal mixing on the geometry of a stripe, as in \cite{verm2, heff}, with that of a single drop. We found similarities between both regimes, such as the asymmetric alignment of vortices along the interfaces and the scale of the vortices being set by the mix-norm index. However, we also found discrepancies between the two geometries. The stripe geometry underwent much more stirring than the disc geometry, though both homogenized reasonably well at similar times. We also varied the Schmidt number and so consequently, as suggested by \cite{heff}, we had $Pe\neq Re$ in this problem. It would appear from this study that the Reynolds number controls the scale whereas the P\'eclet number controls the shape and symmetry of the vortices.

In the second problem (with $Pe=Re$), we examined the effect on mixing dynamics of a lattice of many drops distributed throughout the computational domain. The motivation for this problem was to determine whether the vortex alignment structures observed so far in \cite{foures,marc,verm1,heff} scale as $Pe$ is increased accordingly with the number of drops present in the initial scalar. Indeed, this turns out to be the case as one approaches the turbulent regime but with the slight difference of smaller scales being introduced to perturb the $n = 1$ case but ultimately leading to the same types of dynamics. \color{black} While the structure of the vortices does indeed change as $n$ is increased, we did not find a critical scale at which the alignment structure breaks down. \color{black}   Future studies could look at increasing drops and $Pe$ until such a scale is found.

Looking to future directions,  the stripe and disc geometries appear to provide bounds for the stirring action and it may be of interest to investigate other structures which may or may not sharpen these bounds.
Also, as we found different behaviour for the drop lattices, particularly for the $n=4$ case, it is clear that careful consideration needs to be paid to ensure that small-scale processes are accessible, particularly at higher P\'eclet number.

\vskip6pt

\textbf{Contributions:} All authors contributed equally to the present article.

\textbf{Interests:} We declare we have no competing interests.

\textbf{Funding:} This work was done as part of a PhD project made possible by the National University of Ireland's Travelling Studentship in the Sciences and their support is gratefully acknowledged by the authors.

\textbf{Acknowledgements:} The authors wish to thank the two anonymous reviewers whose comments have greatly improved the presentation and quality of this manuscript.

\textbf{Data Accessibility:} Accompanying animations for the plots may be found here: \color{blue}    https://data.mendeley.com/datasets/n6ctnkmndx/draft?a=06be2dfb-662f-4e9b-96fb-a2f081a5ca8e
\color{black}

Source code is based on a modified version of the solver found in \cite{heff}.


 \bibliographystyle{elsarticle-num}
 \bibliography{cas-refs}












\end{document}